# Peculiarities in the properties of some rare-earth compounds with orthorhombic structures

V. LOVCHINOV, A. APOSTOLOV[a], D. DIMITROV, I. RADULOV, Ph. VANDERBEMDEN[b]
*Institute of Solid State Physics, Bulgarian Academy of Sciences, 72 Tzarigradsko Chaussee Blvd., 1784 Sofia, Bulgaria.*
[a] *University of Sofia, Faculty of Physics, 5 J.Bourchier Blvd.,1164 Sofia, Bulgaria*
[b] *SUPRATECS, Universite de Liege, Institut d'Electricite Montefiore, B28 Sart Tilman, B-6000 Liege, Belgium*

***Corresponding author's e-mail:*** dimitrov@issp.bas.bg

The structural, magnetic, magnetoelectric, and ferroelectric properties of a series of monocrystals with perovskite structures have been examined. The investigations were carried out in the temperature range of 2–800 K and at magnetic fields up to 14 T. The existence of giant magnetoresistance (GMR) for some samples, a giant magnetostiction effect for others and the presence of multiphase ferroelectric states were demonstrated. Various possibilities for practical applications are discussed.

*Keywords*: single crystal manganites; magnetoelectric and ferroelectric properties; multiferroics.

## 1. Introduction

Rare-earth manganites are fascinating, because they display a wide variety of fundamental properties from magnetism to ferroelectricity, from colossal magnetoresistance to semi-metallicity, and because they can be used in a number of important technological applications such as controlling a magnetic memory by an electric field or vice versa, new types of attenuators or transducers etc.

In this paper, we present our investigation on monocrystal samples with an orthorhombic structure, grown in two different space groups: $D_{2h}(16)$ for $La_{0.78}Pb_{0.22}MnO_3$ and $Pr_{0.7}Sr_{0.3}MnO_3$ and $D_{2h}(9)$ for $HoMn_2O_5$ and $TbMn_2O_5$.

The doped perovskite manganites $Ln_{1-x}A_xMnO_3$ (where Ln is a rare-earth ion and A is a divalent ion) from the group $D_{2h}(16)$, which crystallized in different modifications of the perovskite structure, characterized by the parameter deformation of the type $c/\sqrt{2}<b<a$. Many properties of these compounds (especially the giant magnetoresistance GMR, being interesting for practical applications) depend strongly on the carrier density, on the specific zone structure, on the type and the quantity of dopants, on the defects of the crystal and their magnetic structure, or on the applied magnetic fields.

First two compounds of this investigation: $La_{0.78}Pb_{0.22}MnO_3$ and $Pr_{0.7}Sr_{0.3}MnO_3$, are doped by divalent lead and strontium ions. These ions possess ion radii bigger than that of La, and thus they change the deformation of the perovskite lattice. Besides, they cause the appearance of $Mn^{4+}$ and in this way introduce ferromagnetic interactions in the lattice (the interaction $Mn^{3+}$ - $O^{2-}$ $Mn^{4+}$ is positive) Furthermore, they create hole conductivity and an increase in the mobility of d – electrons [1], changing in this way the carrier density.

The second part of this investigation concerning $HoMn_2O_5$ and $TbMn_2O_5$ compounds is aimed to revealing the mediating role of the lanthanide in the appearance of the "giant" magnetostriction effect and electrical polarization.

## 2. Results and discussion

Magnetic and transport properties (Hall effect, electric resistance) of a $La_{0.78}Pb_{0.22}MnO_3$ sample were measured in a wide temperature range (4.2 – 800 K) and magnetic fields up to 14 T, in order to study the effect of the divalent ion. The $La_{0.78}Pb_{0.22}MnO_3$ monocrystal is a typical ferromagnetic material, with $T_c$ = 353 K. Its electrical resistance as a function of the temperature at zero magnetic fields is presented in Fig.1, as an insert. The same figure illustrates the dependence $1/\chi = f(T)$, where $\chi$ is the susceptibility of the sample (right hand curve). As seen from insert of Fig.1 at H = 0 the resistance has a maximal value, where $1/\chi$ tends to zero. The value of the magnetoresistance $\rho(0) - \rho(H)/\rho(H)$ at 300 K is 95 %, and decreases to 45 % at 77 K. The effective magnetic moment of Mn in the paramagnetic region, as determined by our investigation at 4.2K, is $M_{eff} = 5.1\ \mu_B$ (see Fig.1, left hand curve). At 4.2 K, with H parallel or perpendicular to c-axis, the measured values were 4.93 $\mu_B$ and 4.73 $\mu_B$, respectively.

In Fig. 2 experimental data for the Hall voltage as function of the temperature at three different fields are presented. The normal Hall effect coefficient calculated by these data is $R_0 = 1.95.10^{-9}$ $m^3/c$ and the Hall carrier density is $n_H = 3.2.10^{21}$ $cm^{-3}$.

The spontaneous Hall effect coefficient ($R_a = f(T)$) is presented in the same figure, as a dashed line. Course of the curve is negative, and strongly depends on the temperature due to the additional dissipation of the current carriers by the magnetic moments. It is demonstrated by the sharp decrease of the resistance at temperatures lower than 312 K.

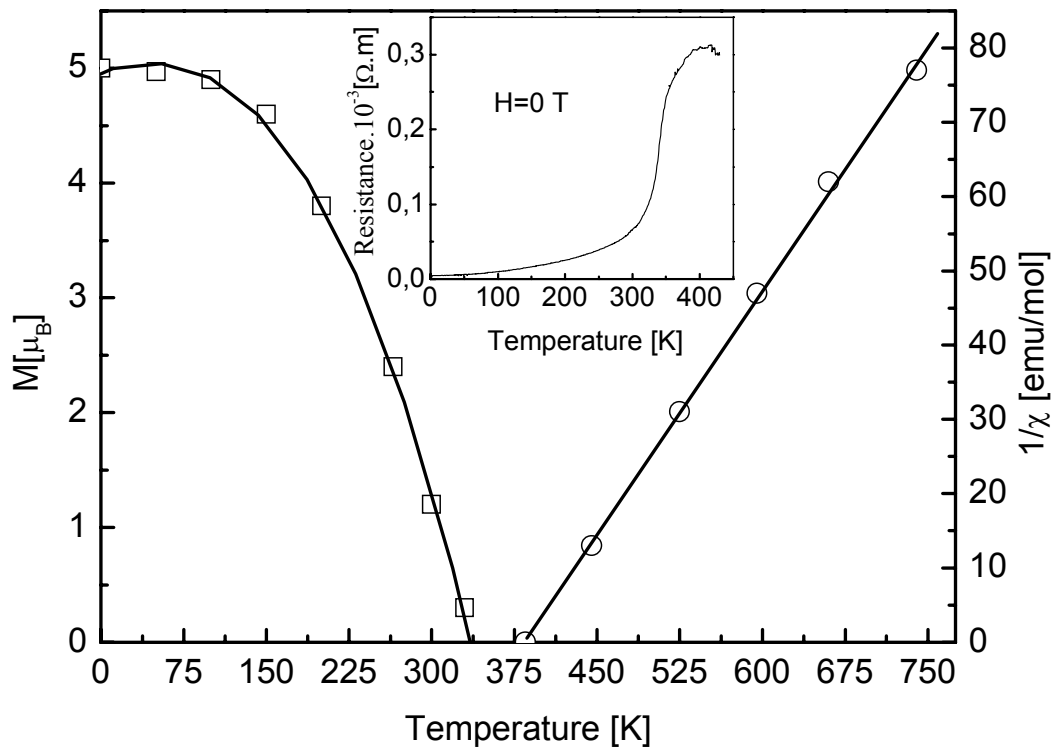

*Fig. 1. Rresistivity (insert), magnetization and reciprocal susceptibility for $La_{0.78}Pb_{0.22}MnO_3$.*

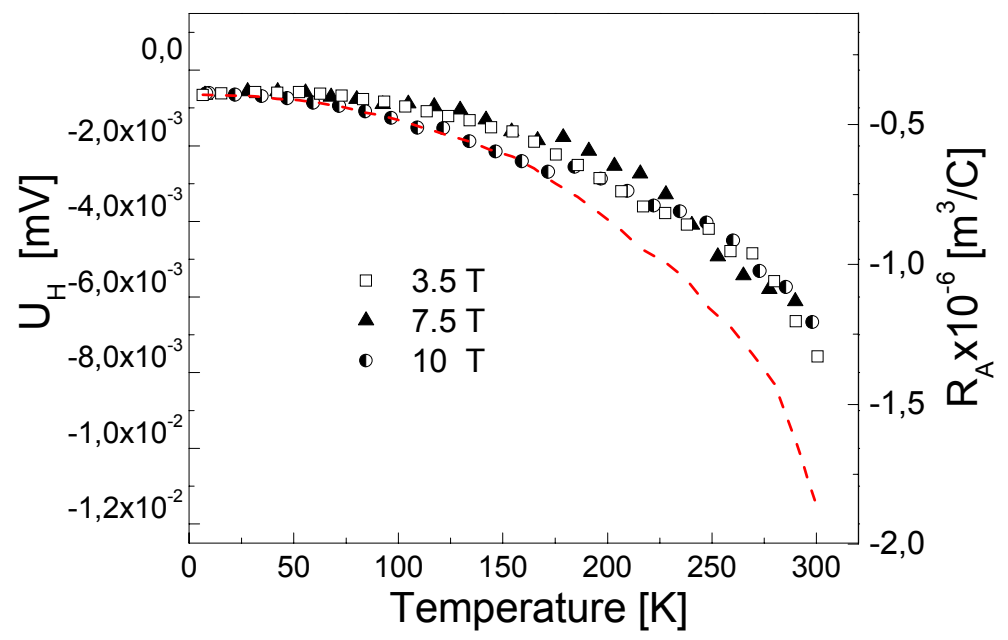

*Fig. 2. Temperature dependence of the Hall voltage and anomalous Hall coefficient.*

The studies carried out indicate that the $La_{0.78}Pb_{0.22}MnO_3$ compound could be useful for the modern microelectronics, since it fulfils two important conditions: it possesses a temperature of magnetic rearrangement (Curie temperature) considerably higher the room temperature and a low electric conductivity, strongly depending on the applied magnetic field.

The investigations on monocrystals of $Pr_{0.7}Sr_{0.3}MnO_3$ were inspired by the supposed simultaneous action of different mechanisms of current carrier dissipation and the expected magnetothermal effect in this compound. The results presented in Fig. 3 reveal the temperature dependence of the resistance and the magnetic susceptibility. It is shown that the monocrystal is paramagnetic above $T_c$ = 270 K and it is ferromagnetic at $T<T_c$. At T = 210 K, there is another phase transition related to the charge arrangement, and, hence, to the lattice deformation (see $\chi = f(T)$). This is exactly the region where the sample is strongly conductive (see Fig. 3 $\rho = f(T)$).

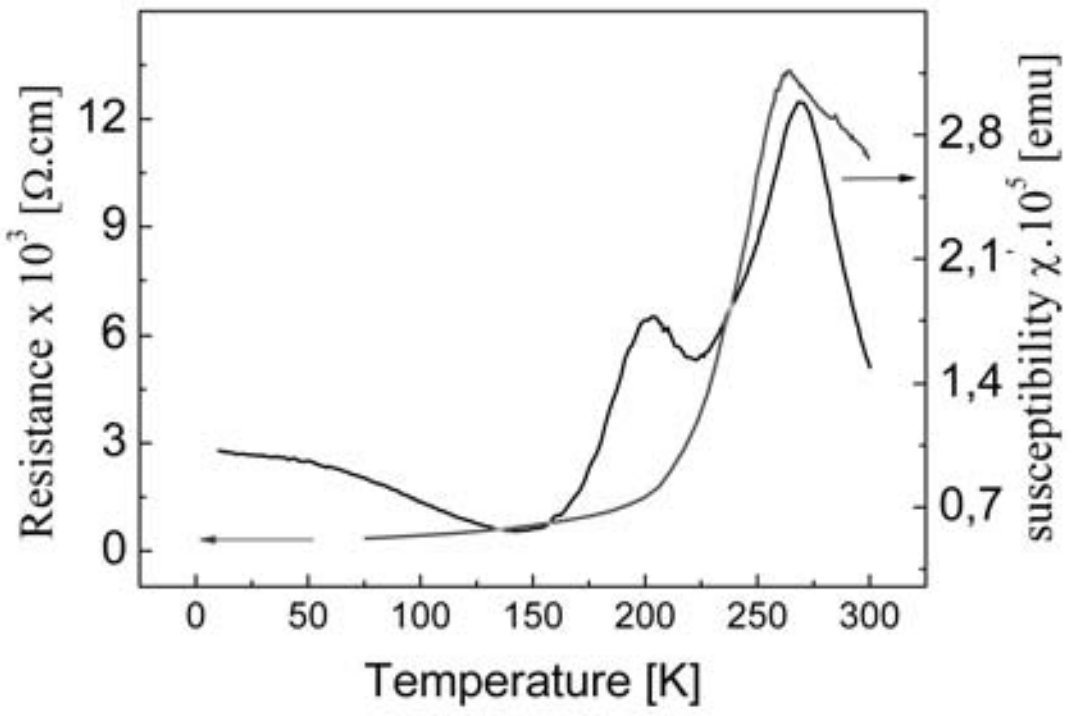

*Fig. 3. Electrical resistivity and magnetic susceptibility for $Pr_{0.7}Sr_{0.3}MnO_3$ monocrystal.*

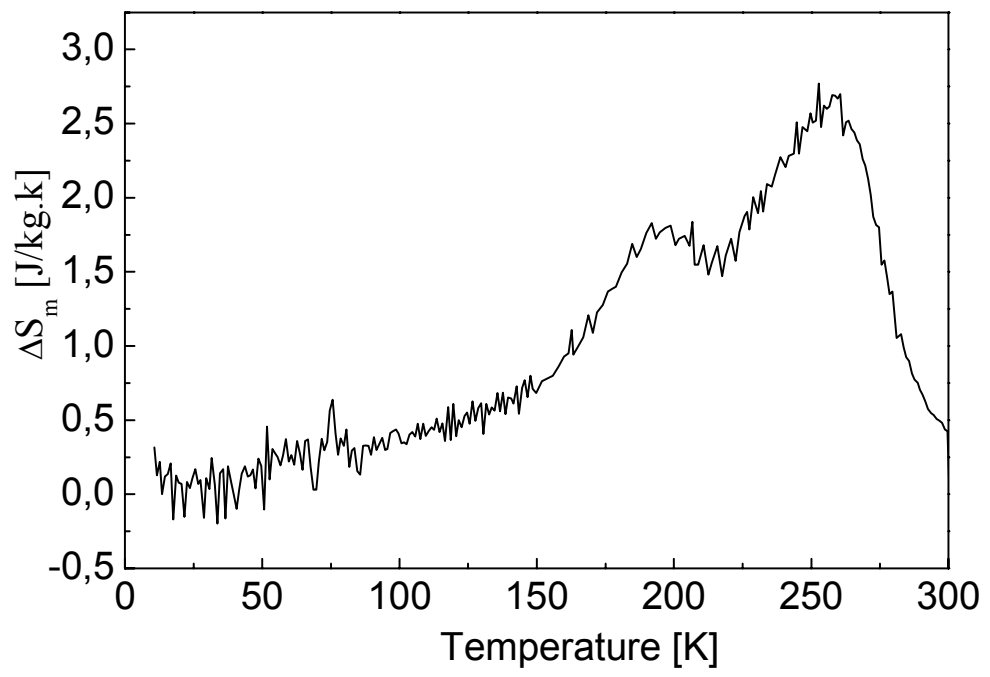

*Fig. 4. The temperature dependence of the entropy change ΔS at H = 1 T.*

In Fig. 4, the entropy change ΔS at H = 1 T depending on the temperature is shown. The values for ΔS were obtained after treatment of the data taking into account the magnetic behavior of the monocrystal. It is seen that the effect is maximal near to the transition point $T_c$ = 270 K. The obtained maximal value of 2.76 J/kg.K indicates that this composition is suitable for application as a substance for magnetic cooling.

Manganites from the space group $D_{2h}(9)$ attract the scientists to study the existing complex magnetoelectric interactions, and provide the opportunity to control them by the application of external magnetic or electric fields [2, 3].

Using methods such as XRD, SEM, and EDAX to study the $HoMn_2O_5$ monocrystal, we have proved that it is orthorhombic with a = 7.333 Å, b = 8.529 Å and c = 5.619 Å. The b-axis of the monocrystal is an axis of easy magnetization and the c – axis – of a difficult one. $Ho^{3+}$, $Mn^{3+}$ and $Mn^{4+}$ occupy respectively the 4g, 4h and 4f locations in the elementary cell.

From the magnetic measurements and the results presented as an insert in Fig.5, one can determine that the monocrystal is paramagnetic above $T_n$ = 44 K with $M_{eff}$=17.4 $\mu_B$ and $\Theta_{paramagnetic}$ is -130 K. Manganese is ordered antiferromagnetically, with a weak ferromagnetism, and Ho remains paramagnetic down to 5 K polarized by this ferromagnetism [4].

When a parallel to b-axis field is applied at 5-40 K, significant hysteresis in the magnetization curves is observed, which decreases with decreasing temperature but remains still significant at 4.2 K ( Fig.5).

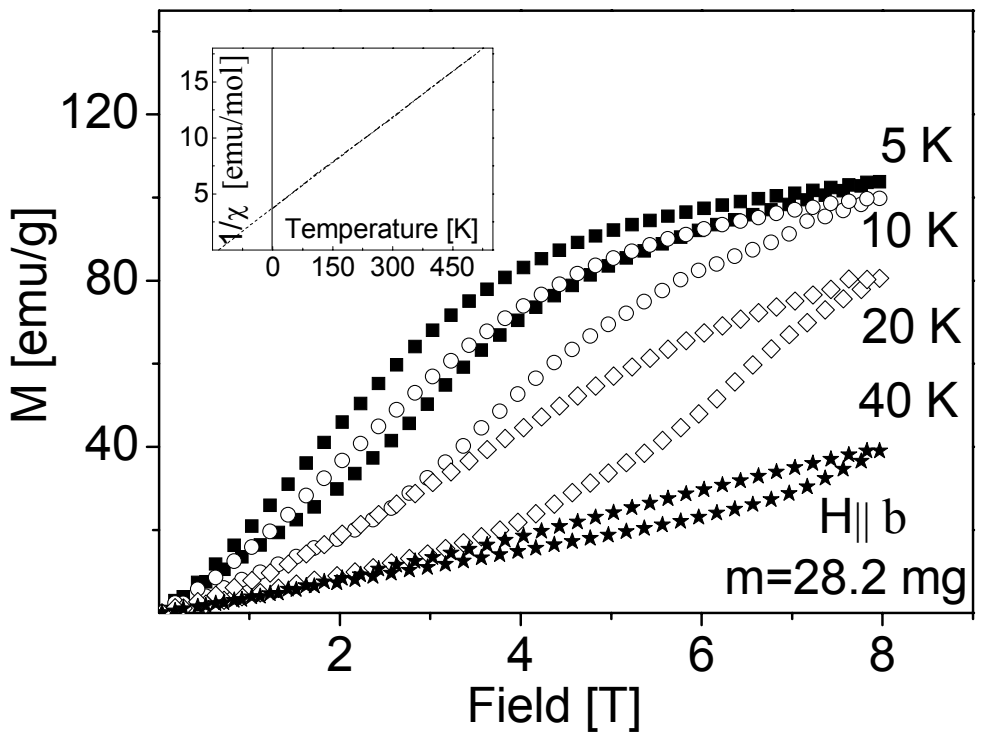

*Fig. 5. Reciprocal susceptibility (insert) and magnetization at different temperatures for $HoMn_2O_5$.*

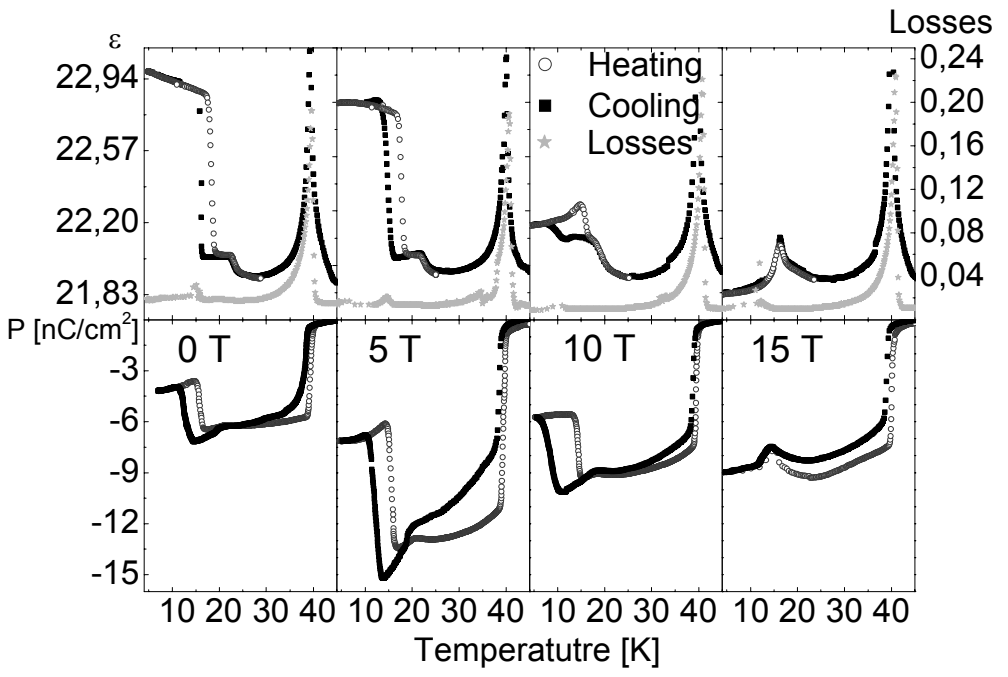

*Fig. 6. Dielectric permeability, dielectric losses and polarization of $HoMn_2O_5$.*

In Fig. 6, measurements of the dielectric permeability, dielectric losses and the polarization of a $HoMn_2O_5$ monocrystal, both with and without magnetic fields up to 15 T, are presented. As seen, by the experimental curves all parameters measured indicate peculiarities at 40 – 42 K, at 20 – 24 K and at 12 – 15 K.

Changes of the antiferromagnetic structure of $HoMn_2O_5$ at H || b and T closed to 20 – 22 K occur due dominantly to $Mn^{4+}$, as also indicated by other authors [5, 6]. Nearly to 12 K, a process of arrangement of $Ho^{3+}$ starts, and is finally completed at lower temperatures.

Fig. 7 presents the experimental data from magnetostriction measurements $\lambda = \Delta L/L$ (where L is the sample's length and ΔL is its elongation) of the $HoMn_2O_5$ monocrystal as a function of the field at different temperatures. At 4.2 K and fields above 1.5 T, effect of "giant" magneto-striction is observed, which reaches a value of $2.10^{-3}$ for a field of 2 T. For comparison, the same value of λ for pure Ho is reached at three times stronger fields of 6 T at 4.2 K. With increasing temperature, this effect appears at stronger fields, with a decrease in λ. The reason for the observed "giant" magnetostriction is the process of overlapping of the exchangeable magnetostriction of Mn with the significant mono-ionic magnetostriction of Ho. The drift of Ho in a magnetic field, (i.e. its magnetostriction), causes the drift of the rest of the ions, despite the fact that Ho is in a non-ordered state. This process is revealed in a cascade of phase transitions.

The conclusion for the importance of the lanthanide in the appearance of the "giant" magnetostriction was checked by carrying out of measurements of the magnetization and magnetostriction of a monocrystal of $TbMn_2O_5$ in magnetic fields up to 15 T and temperatures down to 4 K.

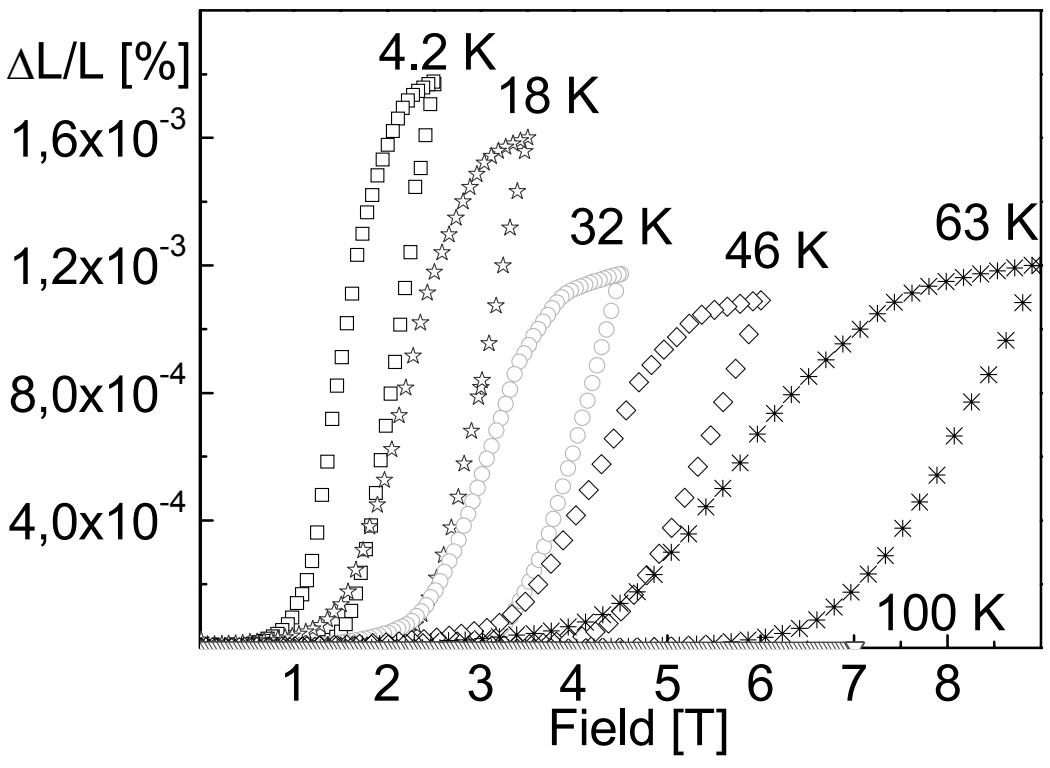

*Fig. 7. Magnetostriction of the $HoMn_2O_5$ monocrystal at different temperatures.*

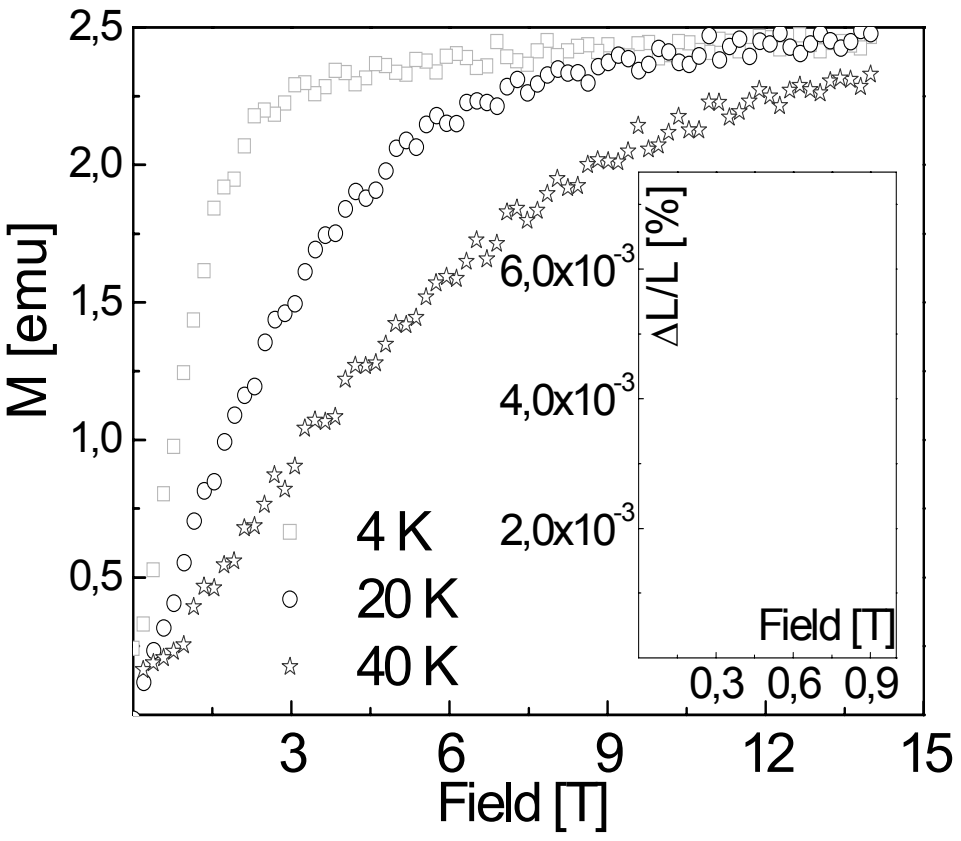

*Fig. 8. Magnetization and magnetostriction (insert) as a function of the magnetic field for $TbMn_2O_5$.*

As seen in Fig. 8, there are some differences both in the types of the M = f(H) curves and in the points of the phase transitions. As compared to the previous Fig 5 concerning for the $HoMn_2O_5$ monocrystal. This behavior of the two compounds is also observed by other authors [3, 6, 7]. The terbium manganite also reveals the effect of "giant" magnetostriction (insert of Fig. 8). It should be emphasize that the value of $\lambda = 6.8.10^{-3}$ for $TbMn_2O_5$ is more than 3 times higher than that $HoMn_2O_5$. On the other hand, this effect for $TbMn_2O_5$ appears at 0.5 T (4.2 K) while for the $HoMn_2O_5$ the effect starts at 1.5 T (4.2 K).

## 3. Conclusions

It has to be concluded that both the manganites from the space group $D_{2h}(16)$ and those of $D_{2h}(9)$ reveal strong magnetoelectic interactions, and due to them a number of interesting effects appear, such as "giant" magnetic resistance, a significant magnetothermal effect and "giant" magnetostriction. The number of the well-known and thoroughly studied pure monocrystals is not very high for the present [8]. However, the opportunity for magneto-electric control of their different properties assures its future intensive investigation and possible practical application. A key to the breakthrough is believed to be the use of multiferroics (like $HoMn_2O_5$ and $TbMn_2O_5$, as presented here), where the ferroic orders of (anti)ferromagnetism and (anti)ferroelectricity coexist.

## Acknowledgements

This work is supported by the CGRI (*Commissariat Général aux Relations Internationales* – Belgium) and by the Bulgarian Academy of Sciences through a joint collaborative research program. The work of I. Radulov is supported by NATO EAP.RIG. 981824.